\documentstyle[12pt]{article}

\font\bb=msbm10 at 10pt
\font\bbT=msbm10 at 7pt

\font\bfs=cmb10 at 7pt
\font\cal=cmsy10 at 9pt
\font\cals=cmsy10 at 8pt

\font\rms=cmr10 at 7pt

\font\sf=cmss10 at 10pt %san serif

\def\0#1{\mbox{\rm#1}}
\def\1#1{\mbox{\bb#1}}
\def\2#1{\mbox{\bf#1}}
\def\3#1{{\cal #1}}
\def\4#1{\mbox{\cals#1}}
\def\5#1{\mbox{\sf#1}} %%sans serif
\def\6#1{\mbox{\rms #1}}
\def\7#1{\mbox{\bfs #1}}
\def\8#1{{\tilde #1}}
\def\9#1{{\breve #1}}

\def\BEq{\begin{equation}}
\def\EEq{\end{equation}}
\def\BEqA{\begin{eqnarray}}
\def\EEqA{\end{eqnarray}}
\def\BEn{\begin{enumerate}}
\def\EEn{\end{enumerate}}

\def\ga{\gamma}

\def\si{\sigma}
\def\Si{\Sigma}

\def\Mbb{\mbox{\bb M}}

\def\Rbb{\mbox{\bb R}}

\def\tav{\hbox{
\kern-1pt\rule[0pt]{1.5pt}{.8pt}{\kern-3.3pt}
\rule[0pt]{.4pt}{5pt}{\kern-4.4pt}
\rule[5pt]{4.5pt}{.8pt}{\kern-3.45pt}
\rule[0pt]{.4pt}{5.3pt}{\kern-1pt}
}}

\def\Ac{{\cal A}}

\def\irm{{\rm i}}
\def\Irm{{\rm I}}

\def\nrm{{\rm n}}
\def\Orm{{\rm O}}

\def\prm{{\rm p}}

\def\Srm{{\rm S}}

\def\ox{\otimes}

\def\Ox{\bigotimes}

\def\Oplus{\bigoplus}

\def\from{\kern-2pt\leftarrow\kern-2pt}

\def\x{\times}

\def\Bar{{\Big |}}

\def\II{|\kern-1pt |}

\def\lo{\stackrel{<}{{}_\sim}}

\def\Cliff{\mathop{\hbox{\rm Cliff}}\nolimits}

\def\Dim{\mathop{\hbox{\rm Dim}}\nolimits}

\def\End{\mathop{\rm End}\nolimits}

\def\ISO{\mathop{\mbox{\rm ISO}}\nolimits}

\def\SO{\mathop{\mbox{\rm SO}}\nolimits}

\def\Spin{\mathop{\mbox{\rm Spin}}\nolimits}

\def\SU{{\mbox{\rm SU}}}

\def\Bar{\kern5pt{\rule[-2.5pt]{.6pt}{9.5pt}}\kern5pt}

\def\GeV{\mbox{ GeV}}

\def\lmult{{\lfloor\kern-5pt\lfloor}}
\def\rmult{{\rfloor\kern-5pt\rfloor}}

\begin{document}

\title{{\bf  Chronon corrections to the Dirac
equation}}

\author{Andrei A. Galiautdinov and David R. Finkelstein \\
\normalsize {\it School of Physics, Georgia Institute of
Technology, Atlanta, Georgia 30332-0430}}

\date{\today}
\maketitle

\abstract
{The Dirac equation is not semisimple. We therefore regard
it as a contraction of a simpler decontracted theory. The decontracted
theory is necessarily purely algebraic and
non-local. 
In one simple model the algebra is
a Clifford algebra with $6N$ generators.
The quantum imaginary $\hbar i$ is the contraction
of a dynamical variable
whose back-reaction provides the Dirac mass.
The simplified Dirac equation
is exactly Lorentz invariant but its symmetry group
is $\SO(3,3)$, a decontraction of
the Poincar\'e
group,
and it has a slight but fundamental non-locality
beyond that of the usual Dirac equation.
On operational grounds the non-locality 
is $\sim 10^{-25}$
sec in size and the associated mass is about the Higgs mass.

There is a non-standard small but unique spin-orbit
coupling $\sim 1/N$, whose observation would be some evidence
for the simpler theory. 
All the fields of the Standard Model
call for similar non-local simplification.}

\vskip20pt

%%%%%%%%PACS numbers: 12.60.-i, 12.10.-g

%%% \pagestyle{myheadings}
%%% \markright{{\rm A.A. Galiautdinov and D.R. Finkelstein},
%%% Non-local corrections to the Dirac equation}

\section{Introduction}

We begin with basic concepts:

A {\em simple} theory is one with simple (irreducible)
dynamical and symmetry groups. What is not simple or
semi-simple we call {\em compound}\/. A {\em contraction} of
a theory is a deformation of the theory in which some
physical scale parameter, called the {\em simplifier},
approaches a singular limit,  taken to be 0
with no loss of generality. The contraction of a simple
theory is in general compound \cite{Segal51, Inonu52, INONU}.
By {\em simplification} we mean the more creative, non-unique
inverse process, finding a simple theory that contracts to a
given compound theory and agrees better with experiment.
The main revolutions in physics of the twentieth century were
simplifications with simplifiers $c, G, \hbar$.

One sign of a compound theory is a
breakdown of reciprocity,
the principle
that every coupling works both ways.
The classic example is
Galilean relativity.
There reciprocity  between space and time
breaks down;
boosts couple time into space and there is no
reciprocal coupling.
Special relativity established reciprocity
by  replacing the compound Galilean bundle of space
fibers over the time base by the simple Minkowski space-time.
Had Galileo insisted on simplicity and reciprocity he could have formulated
special relativity in the 17th century
(unless he were to choose $\SO(4)$ instead of
$\SO(1,3)$).
Every bundle theory violates reciprocity
as much as Galileo's. The bundle group couples
the base to the fiber but not conversely.
Every bundle theory cries out for simplification.

This now requires us to establish
reciprocity
between space-time (base coordinates) $x^{\mu}$
and energy-momenta (fiber coordinates) $p_{\mu}$. [Segal
\cite{Segal51} postulated $x\leftrightarrow p$ symmetry exactly
on grounds of algebraic simplicity;
his work stimulated that of In\"on\"u and Wigner, and ours.
Born \cite{BORN} postulated $x\leftrightarrow p$ reciprocity,
on the grounds
that it is impossible in principle
to measure the usual four-dimensional interval
of two events within an atom.
We see no law
against  measuring space-time coordinates and
intervals at that gross scale.
We use his term ``reciprocity'' in a broader sense
that includes his.]

Einstein's gravity theory and the Standard Model
of the other forces are bundle theories,
with field space as fiber and space-time as base.
Therefore these theories are
ripe for simplification \cite{fs}.
Here we simplify a spinor theory,
guided by criteria of
experimental adequacy, operationality, causality,  and
finity.

Classical field theory is but a singular limit of quantum
field theory; it suffices to simplify the quantum field
theory. Quantum field theory in turn we regard as
many-quantum theory. Its field variables all arise from spin
variables of single quanta. By{\it
quantification} we mean the transition from the one-body to the many-body
theory,  converting yes-or-no predicates about an individual
into how-many predicates about an aggregate of isomorphic
individuals;
as distinct from quantization.
For example, a spinor field theory
arises by quantifying the theory of a single quantum of spin $1/2$.

To unify field with space-time in quantum field theory,
it suffices to unify spin with space-time
in the one-quantum theory, and to quantify
the resulting theory. We unify in this paper
and quantify in a sequel.

Some unification programs concern themselves with simplifying
just  the
internal symmetry group of the elementary particles,
ignoring the fracture between
the internal and external variables.
They attempt to unify (say) the hypercharge, isospin
and color variables,  separate from the space-time
variables.
Here we close the greatest wound first,
expecting that the
internal variables will unite with each other when
they unite with the external variables;
as uniting space with time
incidentally unified the electric and magnetic fields.
We represent space-time variables
$x^{\mu}$
and $p_{\mu}$ as
approximate descriptions of many spin variables,
in one quantum-spin-space-time structure
described in a higher-dimensional spin algebra.
This relativizes the split between field and space-time,
as Einstein relativized the split between space and time.

The resulting quantum atomistic space-time
consists of many small
exactly Lorentz-invariant isomorphic
quantum bits, qubits which we call {\em
chronons}\/. [Feynman, Penrose and Weizs\"acker
attempted to  atomize space or space-time into
quantum spins.
Feynman wrote a space-time vector as the sum of a great
many Dirac spin-operator vectors \cite{FEYNMAN},
$
x^{\mu}\sim \sum_n\gamma^{\mu}(n),
$
Penrose
dissected the sphere $S^2$ into a spin network
\cite{PENROSE}; his work
inspired this program.
Weizs\"acker \cite{WEIZSAEKER},
attempted a cosmology of spin-1/2 urs.
The respective groups
are Feynman's $\SO(3,1)$, Penrose's $\SO(3)$,
Weizs\"ackers $\SU(2)$ and our
$\SO(3N, 3N)$ ($N\gg 1$).]

Simplifying a physical theory generally detaches us
from a supporting condensate.[For Galileo and Kepler,
the condensate was the Earth's crust, and to detach from it they
moved in thought to
a ship or the moon, respectively
\cite{GALILEO, KEPLERSOMNIUM}.]
In the present situation of physics the prime condensate is
the ambient vacuum.
Atomizing space-time enables us to present the vacuum as a
condensate of a simple system,
and to detach from it in thought by a phase transition,
a space-time melt-down.

Chronons carry a fundamental time-unit
$\chi$, one of our simplifiers.
   We have argued that $\chi$ is much greater than the Planck
time and is on the order of the
Higgs time $\hbar/M_H c^2$. [In
an earlier effort to dissect
space-time,  assuming multiple Fermi-Dirac
statistics for the elements \cite{FDR69, DRF96}.
This false start  led  us eventually to the Clifford-Wilczek
statistics
\cite{Wilczek82, NW, wilczek, FINKELSTEINRODRIGUEZ84, FG};
an
example of Clifford-Wilczek statistics is unwittingly
developed in chapter 16 of \cite{DRF96}.]
We now replace the
classical Maxwell-Boltzmann statistics  of space-time events
with the simple Clifford-Wilczek statistics appropriate for
distinguishable isomorphic units.
This  enormously reduces
the problem of forming a theory.

We single out two main quantifications
in field theories like
gravitation and the Standard Model:

A classical quantification assembles a
space-time
from individual space-time points.

A separate quantification constructs a  many-quantum theory
or quantum field theory from a
one-quantum theory
on that space-time.

In the standard physics the space-time quantification tacitly
assumes Maxwell-Boltzmann  statistics for the elements of
space-time, and the field quantification uses Fermi-Dirac or
Bose-Einstein statistics.
The simplified theory we propose uses
one Clifford quantification for all of
these purposes.

In this paper we work only with one-quantum
processes of $N\gg 1$ chronons.
To describe several quanta and their
interactions,
getting closer to field theory and experiment,
will require no further quantification,
but only additional internal combinatory structure
that is readily accomodated within the
one Clifford-Wilczek quantification.

Each physical theory defines at least three algebras
that should be simple:
the associative operator algebra of the system \cite{DRF96,
DRF99},
the
kinematical Lie algebra consisting of possible Hamiltonians, and  the
symmetry Lie algebra of one preferred Hamiltonian.

There is no second quantization.
But there is a second simplification; and a third, and so on,
all of different kinds with different simplifiers.
Each
of the historic revolutions that guide us now
introduced a simplifier, small on the scale of
previous experience
and therefore long overlooked, into the multiplication table and
basis
elements of one or more of these algebras,
and so deformed a
compound algebra into a simpler algebra
that works better.
Among these simplifiers are $c, G$, and $\hbar$.

Here we simplify the free Dirac equation and its underlying
{\em Dirac-Heisenberg} (real unital
associative) algebra
\BEq
\Ac_{\6D \6H} =\Ac_{\6D} \ox \Ac_{\6H},
\EEq
the tensor product of the Dirac and the relativistic
Heisenberg algebras, in turn
defined as follows:

\paragraph{Relativistic Heisenberg algebra}
$\Ac_{\6H}=\Ac [i,\hat p, \hat x]$ is generated
by the imaginary unit $i$ and the space-time and energy-
momentum
translation generators $\hat p_\nu := ip_\nu \equiv - \hbar
{\partial }/{ \partial x^\nu} $ and
$\hat x^\mu :=ix^\mu$, subject to
the relations
\BEqA
\label{eq:HA}
[\hat p^\mu, \hat x^\nu] &=& -i \hbar g^{\mu\nu} , \cr
[\hat p^\mu, \hat p^\nu] &=& 0, \cr
[\hat x^\mu, \hat x^\nu] &=& 0, \cr
[ i, \hat p^\mu] &=& 0, \cr
[ i, \hat x^\mu] &=& 0, \cr
i^2&=&-1.
\EEqA
Here $g^{\mu\nu}$ is the Minkowski metric, held fixed in
this paper. The hats (on $\hat{p}$, for example) indicate
that a factor $i$ has been absorbed to make the operator
anti-Hermitian \cite{adler}.
The algebra $\Ac_{\6H}$ has both the usual associative
product and the Lie commutator product.
As a  real Lie algebra
$\Ac_{\6H}$ is compound, Segal emphasized,
containing the  non-trivial
ideal generated by the unit $i$.

The
orbital Lorentz-group  generators are
\BEq
\hat O^{\mu\nu} :=
iO^{\mu\nu} = -i \left(\hat x^\mu \hat p^\nu - \hat x^\nu
\hat p^\mu
\right).
\EEq
These automatically obey the usual relations
\BEqA
[\hat O^{\mu \nu}, \hat O^{\lambda \kappa}] &=& \hbar \left(
g^{\mu \lambda} \hat O^{\nu \kappa}
- g^{\nu\lambda} \hat O^{\mu \kappa}
- g^{\mu\kappa} \hat O^{\nu\lambda}
+ g^{\nu\kappa} \hat O^{\mu\lambda} \right), \cr
[\hat x^\mu, \hat O^{\nu \lambda}] &=& \hbar \left(
g^{\mu\nu}
\hat x^\lambda - g^{\mu\lambda}
\hat x^\nu \right),\cr
[\hat p^\mu, \hat O^{\nu \lambda}] &=& \hbar
\left( g^{\mu\nu}
\hat p^\lambda -  g^{\mu \lambda}
\hat p^\nu \right),\cr
[ i, \hat  O^{\mu \nu}] &=& 0 .
\EEqA

\paragraph{Dirac algebra}
$\Ac_{\6D}=\Ac[\gamma_{\mu}]$ is generated
by Dirac-Clifford units $\gamma_{\mu}$  subject to the
familiar relations
\BEq
\label{eq:DA}
\{\gamma_\nu , \gamma_\mu \} = 2 g_{\nu \mu}.
\EEq
As usual we write $\gamma_{\mu\nu\dots}$
for the anti-symmetric part
of the tensor $\gamma_{\mu}\gamma_{\nu}\dots$ .

\paragraph{Statistics}
One may define the {\it statistics} of
an (actual, not virtual) aggregate by defining how the
aggregate transforms under permutations of its units.
That is,
to describe $N$ units with given unit
mode space  $V_1$ \cite{DRF96, DRF99} we give, first,
the mode space $V_N$ of the aggregate quantum system
and, second, a simple representation $R_N: \0S_N\to \End
V_N$ of the permutation group $\0S_N$ on the given $N$ units
by linear operators on
$V_N$.
This also defines the quantification
that converts yes-or-no questions
about the individual
into how-many questions about
a crowd.

In {\em Clifford statistics} $\End
V_N$ is a Clifford algebra $C=\Cliff(V_1)$,
and so $V_N$ is a spinor space for that Clifford algebra,
with $C=\End V_N$.
We write $C_1$ for the first-degree
subspace of $C$.
A Clifford statistics is defined by a
projective (double-valued) representation
$R_C: \0S_N\to C_1\subset C$ of the permutation
group $\0S_N$ by first-grade Clifford elements
over the unit mode space $V_1$ \cite{FG}.
To define $R_C$ we associate with the $n$th unit
(for all $n=1, \dots, N$) a Clifford unit $\gamma_n$,
and we represent every swap (transposition
or 2-cycle) $(mn)$ of two distinct units by
the difference $\pm(\gamma_n-\gamma_m)\in C_1$
of the associated Clifford units.

For a free Clifford statistics the units $\gamma_n$ are
independent and the metric $g_{mn}$ is Euclidean.
This representation has dimension $2^{N/2}$ and is reducible.
The irreducible representations have dimension
\cite{Schur1911}
\BEqA
2^{\left\lceil {N-1 \over 2 }\right\rceil}&=&1, 1, 2, 2, 4,
4, 8, 8, 16,\dots\cr
\mbox{\rm for } \Dim V_1 = N&=&0, 1, 2, 3, 4, 5, 6, 7, 8,
\dots \quad .
\EEqA

Some useful terms:

   A {\em cliffordon} is a
quantum with Clifford statistics.

A {\em squadron} is a quantum aggregate of cliffordons.

A {\em sib} is a quantum aggregate of bosons.

A {\em set} of quanta is an
aggregate of fermions.

A {\em sequence} of quanta is a
aggregate of Maxwell-Boltzmann quanta with a given
sequential order \cite{FDR69}.

$R_C$ can be extended to a spinor
representation of $\SO(N)\supset \0S_N$
on the same spinor space $\Si(N)$.

The symmetry group $G_U$ of the quantum kinematics for a universe $U$ of
$N_U$ chronons is an orthogonal group
\BEqA
G_U&=&\SO(N_{U+}, N_{U-}),\cr
N_U&=&N_{U+} +N_{U-}.
\EEqA
The algebra of observables of $U$
is the simple
finite-dimensional real Clifford algebra
\BEq
C_U=\Cliff (V_1) =
\Cliff[1, \gamma(1),\dots, \gamma(N_U)]
\EEq
generated by the  $N_U$ Clifford units
$\gamma(n), \; n= 1, \dots, N_U$ representing
exchanges.
The Clifford units $\gamma(n)$
span a vector space $V_1 \cong C_1$ of first-grade elements
of $C_U$.

Within
$C_U$ we shall construct a simplified Dirac-Heisenberg
algebra
\BEq
\breve{\Ac}_{\6D \6H} =
{\Ac}[\breve{i},
\breve p, \breve x,\tilde \gamma]
\subset
C(V_1)
\EEq
whose commutator Lie algebra is simple
and which
contracts to the usual
Dirac-Heisenberg algebra
$\Ac_{\6D \6H}$
in the continuum limit.
We factor  ${\breve \Ac_{\6D \6H}}$
into the Clifford product
\BEq
\label{eq:CLIFFORDPRODUCT}
\breve \Ac_{\6D \6H}=\Cliff(N{\bf
6_0}) =  \Cliff((N-1){\bf 6_0})\sqcup\Cliff({\bf 6_0})
\EEq
of two Clifford algebras,
an ``internal'' algebra from the last hexad
and an ``external" algebra from
all the others.

We designate our proposed simplifications of $\gamma$ and
$i, \hat p$, $\hat x$, and $\hat
O$  by  $ \tilde \gamma $ and $\breve{i},
\breve p, \breve x$ and $ \breve O$.
In the limit $\chi\to 0$
the
tildes
$\tilde{\phantom{m}}$ disappear
and
the breves $\breve{\phantom{m}}$ become hats $\hat{\phantom{m}}$.

We use the following quadratic spaces:

$N{\1R}:=\1R \oplus \dots \oplus \1R$ (with $N$ terms)
$=\Oplus_1^N \1R$ is the positive-definite $N$-dimensional
real quantum-mode space.

$-N{\1R}$ is the corresponding
negative-definite space.

$\1M$ is Minkowski space with
signature $1-3$.

$-\1M$ is the same space with the opposite
signature $3-1$.

Also,
\BEqA
\label{eq:SPACES}
X\ominus Y&:= & X\oplus (-Y)\cr
   {\bf 1}&:=& \1R ,\cr
{\bf -1}&:=& {-\1R} ,\cr
{\bf 3}&:=& 3\1R ,\cr
{\bf -3}&:=& -3{\1R} ,\cr
\Mbb&:=&{\bf 1}\ominus {\bf 3},\cr
-\Mbb&:=&{\bf 3}\ominus {\bf 1},\cr
{\bf 6_0} &:=& {\bf 3} \ominus {\bf 3}.
\EEqA
$\1M$ and $-\1M$ are tangent spaces to
Minkowski space-times
and support natural
representations of the Lorentz group.

\section{Simplification of the relativistic Heisenberg
algebra}
\label{sec:STVARIABLES}

As already mentioned, field theory employs a
compound field-space-time bundle with space-time for base
and field-space for fiber; just as Galilean space-time
is a four-dimensional bundle
with $\Rbb^3$ for base and  $\Rbb^1$ for fiber.
The prototype is the covector field,
where the fiber is the cotangent space to space-time,
with coordinates that we designate by $p_{\mu}$.

We assume that in experiments of sufficiently high
resolution the
space-time tangent bundle (or the
Dirac-Heisenberg algebra) manifests itself as a simple
quantum-field-space-time synthesis.
The
space-time variables $x^\mu$ and  the tangent space
variables $p_\mu$ unite into one simple construct,
as space and time have already united.
Now, however, the simplification requires an atomization,
because the field variable actually derives from
an atomic spin.

We first split the space-time tangent bundle
into quantum cells. 
The minimum number of elements in a cell
for our simplification is six: four for space-time and two
for a complex or symplectic plane. We provisionally adopt the
hexadic cell. Earlier work, done before our present stringent
simplicity standard, assumed a pentadic cell
\cite{FINKELSTEINRODRIGUEZ84}. This provided no natural
correspondent for the energy-momentum operators.

$N$ hexads define a unit mode space $V_1 = N {\bf 6_0}$.
Each term ${\bf 6_0}$ has a Clifford algebra $\Cliff(3,3)$
whose spinors have eight real components, forming an
$\28_0$. [Eight-component spinors have also been
used in physics by Penrose \cite{PENROSE}, Robson and
Staudte \cite{ROBSON}, and Lunsford \cite{LUNSFORD};
though not to unify spin with
space-time.] The
spinors of $V$ form the spinor space $$\Ox_1^N
\28_0=\28_0^N$$.

We do not deal with empty space-time.
We explore space-time with one relativistic quantum
spin-$1\over2$ probe of rest mass  $m \sim 1/\chi$.
We express the usual spin operators $\ga^{\mu}$, space-time position operators
$x^{\mu}$, and energy-momentum  operators $p_{\mu}$
of this probe
as contractions of operators
in the Clifford algebra $
\Cliff(3N,3N)$.

We write the dynamics of the usual contracted, compound
Dirac
theory in manifestly covariant form,
with a Poincar\'e-scalar Dirac operator
\BEq
\label{eq:DIRACDYNAMICS}
D=\ga^{\mu} p_{\mu} - mc .
\EEq
$D$ belongs to the algebra of operators
on spinor-valued functions $\psi(x^\mu)$ on space-time.
Any physical spinor
$\psi(x^{\mu})$  is to obey the dynamical equation
\BEq
\label{eq:DYNAMICS}
D\psi=0.
\EEq
We simplify the dynamical operator $D$,
preserving the form of the dynamical equation
(\ref{eq:DYNAMICS}).

The compound symmetry group for the Dirac equation
is the covering group of the Poincar\'e group
$\Irm{\Srm}\Orm(\Mbb)$.
We represent this as the contraction of a
simple group $\SO(3,3)$ acting on the
spinor pseudo-Hilbert (ket) space of
$6N$ Clifford generators
$\ga^{\omega}(n)$ ($\omega =0, \dots, 5$; $n=1,\dots, N$)
of the orthogonal group $\SO(3N, 3N)$.
The size of the
experiment fixes the parameter $N$.

We first simplify the anti-Hermitian space-time and
energy-momentum translation generators
$\hat p_\mu$ and $\hat x^\nu$, not the associated Hermitian
observables $p_\mu$, $x^\nu$.
Then we simplify the Hermitian operators
by multiplying the anti-Hermitian ones by a
suitably simplified $i$ and symmetrizing the product.

As in Dirac one-electron theory, we use the
spinor representation $\Srm\prm\irm\nrm(\Mbb)$ of
$\Srm\Orm(\Mbb)$ to describe the contracted generators
$\hat S^{\mu\nu}$ of rotations and
boosts.
The spin generators  are represented by second-degree elements
\BEq
\label{eq:DIRACGENERATORS}
\hat S^{\mu \nu}:=
\frac{\hbar}{4}\left[\gamma^\mu, \gamma^\nu\right]
\equiv
\frac{\hbar}{2}\gamma^{\mu\nu}, \quad
\mu, \nu = 0,
\dots, 3
\EEq
of the Clifford algebra $\Cliff(1,3)$.

We simplify
$\ISO(\Mbb)$ within
$\Spin(N{\bf 6_0})$ by representing the
simplified space-time
symmetry generators of the probe
by second degree elements of
$\Cliff V_1$.
We associate the position and momentum axes
with the $\ga^4$ and $\ga^5$ elements of the hexad respectively,
so that
an infinitesimal orthogonal transformation
in the 45-plane couples momentum into position.
This accounts for the symplectic symmetry of
classical mechanics and the $i$ of quantum mechanics.

We therefore define the simplified $\breve
i$, $\breve x^\mu$, and
$\breve p^\nu$ of the probe  by
\BEqA
\breve i \equiv \frac{1}{N-1}\sum_{n=1}^{N-1}
\breve i(n) &:=& \frac{1}{N-1} \sum_{n=1}^{N-1}
\gamma^{45}(n) ,\cr\cr
\breve x^\mu \equiv \sum_{n=1}^{N-1} \breve
x^{\mu}(n) &:=& -\chi
\sum_{n=1}^{N-1}
\gamma^{\mu 4}(n) , \cr\cr
\breve p^\nu
\equiv \sum_{n=1}^{N-1} \breve
p^{\nu}(n) &:=& \phi
\sum_{n=1}^{N-1}
\gamma^{\nu 5}(n),
\EEqA
where $\chi$, $\phi$ and $N$ are simplifiers of our theory,
and
\BEq
\gamma^{\rho \sigma}(n)
:= \frac{1}{2} [\gamma^{\rho}(n), \gamma^{\sigma}(n)].
\EEq

To support this choice for the expanded generators we form
the following commutation relations among them ({\it
cf.} \cite{SNYDER, Segal51,DOPLICHER}):
\BEqA
\label{eq:EXTERNALCOMMRELSINGLE}
[\breve p^\mu, \breve x^\nu] &=& -2 \phi
\chi (N-1)\;  g^{\mu\nu} \, \breve i,
\cr \cr
[\breve p^\mu, \breve p^\nu] &=& -\frac{4
\phi^2}{\hbar}
\;   \breve L^{\mu \nu} , \cr
\cr [\breve x^\mu, \breve x^\nu] &=& -\frac{4
\chi^2}{\hbar}\;
     \breve L^{\mu \nu},
\cr \cr
[\breve i, \breve p^\mu ] &=&
- \frac{2\phi}{\chi (N-1)}
\;   \breve x^\nu, \cr \cr
[\breve i, \breve x^\mu ] &=&
+ \frac{2\chi}{\phi (N-1)}
\;   \breve p^\mu.
\EEqA

   In (\ref{eq:EXTERNALCOMMRELSINGLE}),
\BEqA
\label{eq:ANGULARMOMENTA}
\breve L^{\mu\nu}&:= &\frac{\hbar}{2}
\sum_{n=1}^{N-1} \gamma^{\mu\nu}(n) ,\cr
\breve J^{\mu\nu}&:= &\frac{\hbar}{2}
\sum_{n=1}^{N} \gamma^{\mu\nu}(n)
\equiv L^{\mu\nu}+S^{\mu\nu} .
\EEqA
where $\breve S^{\mu\nu}$ is
the Dirac spin operator ({\it cf.}
(\ref{eq:DIRACSPINOPERATOR})),

(\ref{eq:EXTERNALCOMMRELSINGLE}) 
incorporates two decontractions:
one 
leading to finite
commutators between coordinates
of the Snyder type,
and one 
leading to finite commutators
between $\hbar i$ and the coordinates and momenta
of the Segal type.
Both are necessary for simplicity.

The Snyder decontraction makes the theory more 
non-local than the Dirac equation.
In the contracted theory,
the coordinates $x, y, z$ commute.
This means that in principle
one can produce the single quantum
at a definite place
and register it at a definite place.
To be sure,
to do so will mix positive and
negative energy levels.
In the more physical many-quantum theory,
a pair will be created
in these processes.
Nevertheless, in the standard
interpretation of the quantum theory it is still
possible in principle
to precisely determine the operators $x,y,z$
with arbitrary precision at one instant,
before the pair separates.

In the decontracted theory,
any one of the operators $x, y, z$
can be determined
with arbitrary precision,
say $z$.
Its spectrum will then be discrete.
The operators $x,y$
will then have fundamental 
indeterminacies,
depending on the magnitude
of $L_z$
and the constant $\chi$.
Thus the single quantum 
can no longer be localized in principle.
This non-locality is
intrinsic to 
the space-time-momentum-energy-spin
unification that  makes the theory
simpler.

$J^{\mu\nu}$
obeys the Lorentz-group commutation relations:
\BEq
[\breve J^{\mu \nu}, \breve J^{\lambda \kappa}] = \hbar
\left( g^{\mu \lambda} \breve J^{\nu \kappa}
- g^{\nu\lambda} \breve J^{\mu \kappa}
- g^{\mu\kappa} \breve J^{\nu\lambda}
+ g^{\nu\kappa} \breve J^{\mu\lambda} \right),
\EEq
and generates a total Lorentz transformation of the variables $x^{\mu}$,
$p_{\mu}$,
$i$ and $S^{\mu\nu}$:
\BEqA
[\breve x^\mu, \breve J^{\nu \lambda}] &=& \hbar \left(
g^{\mu\nu}
\breve x^\lambda - g^{\mu\lambda}
\breve x^\nu \right),\cr
[\breve p^\mu, \breve J^{\nu \lambda}] &=& \hbar
\left( g^{\mu\nu}
\breve p^\lambda -  g^{\mu \lambda}
\breve p^\nu \right),\cr
[ \breve i, \breve  J^{\mu \nu}] &=& 0 ,\cr
[ \breve S^{\mu\nu}, \breve  J^{\lambda \kappa}] &=& 0.
\EEqA

There is a mock orbital
angular momentum generator of familiar appearance,
\BEq
\label{eq:MOCKORBITAL}
\breve O^{\mu\nu} := -\breve i \left( \breve
x^\mu
\breve p^{\nu} - \breve x^\nu \breve
p^{\mu}\right).
\EEq
$\breve O$ too obeys the Lorent group commutation relations.
We relate $\breve L^{\mu \nu}$ and $\breve
O^{\mu\nu}$ in
Sec.\ref{sec:LORENTZGENERATORS}.

Since the usual complex unit $i$ is central and the
simplified $\breve i$ is not, we suppose that the contraction
process includes a projection that restricts the probe to
one of the two-dimensional invariant subspaces of $\breve
i$,  associated with the maximum negative eigenvalue $-1$ of
${\breve i}^2$.  This represents a condensation that aligns
all the mutually commuting hexad spins
$\gamma^{45}(n)$ with each other,  so that
\BEq
\label{eq:CONTRLIMIT}
\gamma^{45}(n)\gamma^{45}(n') \longrightarrow -1,
\EEq
for any $n$ and $n'$.
We call this the condensation of $i$.

Projection onto a sharp value of $i$ kills
$i$-changing variables like $x^{\mu}$ and $p_{\mu}$.
Only $\SO(2)$-invariant combinations like $\chi^2 p^2 +\phi^2 x^2$ should
survive.
Nonetheless one observes position and momentum
separately.
This is a spontaneous symmetry-breaking
by the vacuum condensate,
analogous to the fact that
a crystal in its ground state,
with spherically symmetric Hamiltonian, can have a non-zero
internal magnetization.

Under Wigner time-reversal,
$t\to -t$ and $i\to -i$.
Since the variable $t$ is chosen by the experimenter,
not the system,
we must suppose that $i$
too is mainly fixed by the experimenter, not the system.
But since the boundary between system and experimenter
   is somewhat
arbitrary, we must therefore
suppose that the entire universe
contributes uniformly to $i$;
it is simply that the system is much smaller
than the experimenter,
and influences $i$ less.
This fits with an earlier theory of $i$
as a St\"uckelberg-Higgs variable
that imparts mass to some
otherwise massless gauge vector bosons
\cite{STUECKELBERG1938,
FINKELSTEINJAUCH1959, FINKELSTEINJAUCH1962, HIGGS}

Then the  momentum variables $p_\mu$
that we usually attribute to the system,
for example,
are actually $i$-invariant bilinear combinations
$P_{[\mu]}^{\rho\si}J_{\rho\si}$ of experimenter standards
$P_{[\mu]}^{\rho\si}$ and the system tensor $J_{\rho\si}$.
As creatures of the space-time condensate
we do not experience the symmetry of the dynamics
that produced it,
but only its residual symmetries.
The spontaneously broken symmetries reappear
when the condensate melts down.

To recover the canonical commutation relations for
$\breve x^{\mu}$ and $\breve p_{\mu}$ we must impose
\BEq
\chi\phi (N-1) \equiv {\hbar\over 2}
\EEq
and assume that
\BEqA
\label{eq:CONTRLIMIT1}
\chi &\rightarrow & 0, \cr
\phi &\rightarrow & 0, \cr
N&\to& \infty .
\EEqA
Then the relations (\ref{eq:EXTERNALCOMMRELSINGLE}) reduce to
the commutation relations (\ref{eq:HA}) of the relativistic
Heisenberg algebra $\3A_H$ as required.

The three parameters $\chi, \phi, 1/N$ are subject to one
constraint $\chi\phi (N-1) = \hbar/2$ leaving two independent
simplifiers. $N$ is not a physical constant like
$\hbar$ and $c$, but depends on the scope of the experiment,
and is under the experimenter's control. $N$-dependent
effects might appear as curious boundary effects.
   We set a cosmological limit
$N\lo N_{\6M\6a\6x}$ below.

This leaves one $N$-independent physical constant with the
dimensions of time.
We can consider two contractions,  $\chi\to 0$
with $N$ constant,  and $N\to \infty$
with $\chi$ constant.
They combine into
the  continuum limit $\chi\to 0$, $N\to \infty$.
We  fix one simplifier $\chi$ in Sec.\ref{sec:DIRACDYNAMICS}
by supposing that the mass of a probe approaches a finite
limit as
$N\to\infty$.

\section{Orbital, spin, and total angular momentum}
\label{sec:LORENTZGENERATORS}

As was shown in Sec.\ref{sec:STVARIABLES}, three sets of
operators obeying Lorentz-group commutation relations appear
in our theory.
$\breve  L^{\mu\nu}$ represents the simpified {\em orbital angular
momentum}
generators, $\breve S^{\mu\nu}$ represents the {\em spin} angular
momentum, and
$\breve J^{\mu
\nu}$ represents the simplified {\em total angular momentum} generators.
There is a mock orbital angular momentum $\breve O^{\mu\nu}$
(\ref{eq:MOCKORBITAL}).

In this section we show that $\hat O\to \hat L$  in
the contraction limit.

Consider $\breve O^{\mu\nu}$. By definition,
\BEqA
\breve O^{\mu\nu}&=& -\left(\breve
x^\mu \breve p^{\nu} - \breve x^\nu \breve
p^{\mu}\right) \breve i \cr
&=&  +\frac{\chi \phi}{N-1} \left(
\sum_{n=1}^{N-1}
\gamma^{\mu 4}(n)
\sum_{n'=1}^{N-1}
\gamma^{\nu 5}(n') -
\sum_{n=1}^{N-1}
\gamma^{\nu 4}(n)
\sum_{n'=1}^{N-1}
\gamma^{\mu 5}(n') \right) \;
\sum_{m=1}^{N-1}
\gamma^{45}(m) \cr\cr\cr
&=&  +\frac{\chi \phi}{N-1}
\sum_{n}\left(
\gamma^{\mu 4}(n)
\gamma^{\nu 5}(n) -
\gamma^{\nu 4}(n)
\gamma^{\mu 5}(n) \right) \;
\sum_{m}
\gamma^{45}(m) \cr\cr\cr
&& +\frac{\chi \phi}{N-1}
\sum_{n\neq n'}\left(
\gamma^{\mu 4}(n)
\gamma^{\nu 5}(n') -
\gamma^{\nu 4}(n)
\gamma^{\mu 5}(n')
+\gamma^{\mu 4}(n')
\gamma^{\nu 5}(n) -
\gamma^{\nu 4}(n')
\gamma^{\mu 5}(n)\right) \cr\cr
&& \times
( \gamma^{45}(n) + \gamma^{45}(n'))
\cr\cr\cr
&& +\frac{\chi \phi}{N-1}
\sum_{n\neq n'}\left(
\gamma^{\mu 4}(n)
\gamma^{\nu 5}(n') -
\gamma^{\nu 4}(n)
\gamma^{\mu 5}(n')
+\gamma^{\mu 4}(n')
\gamma^{\nu 5}(n) -
\gamma^{\nu 4}(n')
\gamma^{\mu 5}(n)\right) \cr\cr
&& \times
\sum_{m\neq n, m\neq n'}
\gamma^{45}(m) \cr\cr\cr
&=&  -\frac{2\chi \phi}{N-1}
\sum_{n}
\gamma^{\mu \nu}(n)
\gamma^{45}(n) \;
\sum_{m}
\gamma^{45}(m) \cr\cr\cr
&& +\frac{\chi \phi}{N-1}
\sum_{n\neq n'}\left(
\gamma^{\mu 4}(n)
\gamma^{\nu 5}(n') -
\gamma^{\nu 4}(n)
\gamma^{\mu 5}(n')
+\gamma^{\mu 4}(n')
\gamma^{\nu 5}(n) -
\gamma^{\nu 4}(n')
\gamma^{\mu 5}(n)\right) \cr\cr
&& \times
\sum_{m\neq n, m\neq n'}
\gamma^{45}(m) .
\EEqA

Thus, in the contraction limit
(\ref{eq:CONTRLIMIT})-(\ref{eq:CONTRLIMIT1}) when
condensation singles out the eigenspace of
$\gamma^{45}(n)\gamma^{45}(n')$ with
eigenvalue -1,
\BEq
\hat O^{\mu\nu} \longrightarrow \hat
J^{\mu\nu} - \hat S^{\mu \nu} \equiv \hat L^{\mu \nu},
\EEq
as asserted.

\section{Simplified Dirac dynamics}
\label{sec:DIRACDYNAMICS}

Dirac's {\it one-body} theory in real (Majorana) form uses
the operator algebra $A_{\6D\6H}$
acting on a vector space
\BEq
V_1 := \Sigma^{-\mbox{\bbT M}}.
\EEq
of spinor-valued wavefunctions,
mapping  the space-time
to the spinor space
$\Si=\Sigma(-\Mbb)$
over the Minkowski space-time $-\Mbb$.
This exhibits part of the compound structure
we must simplify by decontraction,
the split between spin space $\Sigma$
and space-time $\Mbb$.
We construct the new space entirely
from spins,
replacing 
the infinite-dimensional
function space $V_1$  by a spinor space of high but finite dimensionality.

To simplify Dirac's spin-${1\over 2}$ dynamics, we
regard the position of the probe as the resultant of $N$
quantum steps, each represented by one hexad of chronons.
We identify the spin variables of the probe with those of the
last hexad in (\ref{eq:CLIFFORDPRODUCT}), the growing tip of
the world line of the probe.

We  thereby simplify the Dirac-Heisenberg algebra
$\Ac_{\6D \6H}$
to ${\breve \Ac_{\6D \6H}} := \Cliff(N{\bf 6_0})$, the
Clifford algebra of a large squadron of cliffordons.

To construct the contraction from ${\breve \Ac_{\6D \6H}}$
to $\Ac_{\6D \6H}$, we group the generators of $\Cliff(3N,
3N)$ into $N$ hexads $\ga^{\omega}(n)$ ($\omega = 0, \dots,
5$; $n=1,\dots, N$). Each hexad algebra acts on
eight-component real spinors in $\28_0$.
Hexad $N$ will be used for the spin of the quantum.
The remaining $N-1$ hexads provide the space-time variables.

We identify the usual Dirac gammas $\ga^{\mu}$ for $\mu=0,
\dots, 3$ of $\Cliff(-\Mbb )$ with second-degree elements
of the last hexad:
\BEq
\gamma^{\mu} \cong \tilde \gamma^{\mu} := \gamma^{\mu 5} (N)
\EEq

Dirac's spin generators $\hat S^{\mu\nu}$
(\ref{eq:DIRACGENERATORS}) simplify to
the corresponding 16 components of the tensor
\BEq
\label{eq:DIRACSPINOPERATOR}
\breve S^{\omega \rho}:= \frac{\hbar}{2}\,
\gamma^{\omega \rho}(N),
\EEq
where $\omega , \rho = 0, \dots , 5$
and $\mu,\nu = 0, \dots, 3$.

It is now straightforward to simplify the
Dirac equation $D\psi=0$ of (\ref{eq:DYNAMICS}).
The internal degrees of freedom will be seen to contribute a
rest mass term $m_{\chi}=\hbar/2\chi c$, and for simplicity
we take this to be the entire rest mass of the Dirac equation,
omitting any bare mass term in $\tilde D$.
We simplify $D\to \tilde D$ and extend the internal symmetry
group from $\SO(1,3)$ to the group $\SO(3,3)$ of a hexad by
setting
\BEq
\label{DYNAMICS}
\tilde{D} := \frac{2\phi}{\hbar^2} \; \breve
S^{\omega\rho}
   \breve L_{\omega \rho},
\EEq
where ({\it cf.} (\ref{eq:ANGULARMOMENTA}))
\BEq
\breve L_{\omega \rho}:= \frac{\hbar}{2}\sum_{n=1}^{N-1}
\gamma_{\omega \rho}(n).
\EEq

Our proposed dynamical operator is invariant under a
conformal group
$\SO(3,3)$
whose contraction includes the Poincar\'e group.
[Our symmetry group $\SO(3,3)$  incorporates and extends the
$\SO(3, 2)$ symmetry possessed by Dirac's dynamics for an
electron in de-Sitter space-time
\cite{DIRAC35}. That dynamics has
the form
$$
D'=\frac{1}{\hbar R}{\hat S}^{\omega \rho} {\hat
O}_{\omega\rho}
- mc,
$$
where ${\hat S}^{\omega \rho}$ and $\hat O_{\omega\rho}$ are
the five-dimensional spinorial and orbital angular momentum
generators and $R$ is the radius of the de-Sitter universe.
Its group is still compound, not simple,
unifying translations, rotations and boosts,
but not symplectic transformations.]

A complete set of commuting generators for the Poincar\'e
group $\ISO(1,3)$ consists
of the time translation generator $\hat p_0$,
the rotation generator
$\hat L_{12}$, and the boost generator
$\hat L_{03}$.
In the present context, we adjoin the imaginary unit
$i$. In the proposed simplification $\ISO(1,3) \x
\SO(2)\from
\SO(3,3)$ these simplify according to
$\hat p_0\from \breve L_{04}$, $\hat L_{12}\from \breve
L_{12}$, $\hat L_{03}\from \breve L_{03}$, $i\from \breve
L_{45}$. A commuting set cannot contain both $\breve L_{04}$
and $\breve L_{45}$. Since varying energy is more familiar
than varying $i$, in a first treatment we hold $\breve
L_{45}$ constant and couple different masses in one
representation.

\section{Reduction to the Poincar\'e group}

We now assume a condensation that reduces $\SO(3,3)$ to
its subgroup
$\SO(1,3)\x\SO(2)$. 
Relative to this reduction, the $\tilde
D$ of (\ref{DYNAMICS}) breaks up into
\BEqA
\tilde{D} &=& \frac{\phi}{2} \, \gamma^{\omega\rho}(N)
\sum_{n} \gamma_{\omega \rho}(n) \quad \quad \quad
\quad
\quad \quad
(\omega, \rho = 0, 1, \dots , 5 ) \cr\cr\cr &=& \phi \,
\gamma^{\mu 5}(N)
\sum_{n} \gamma_{\mu 5}(n) + \phi \,
\gamma^{\mu 4}(N)
\sum_{n} \gamma_{\mu 4}(n) + \phi \,
\gamma^{\mu\nu}(N)
\sum_{n} \gamma_{\mu\nu}(n) \cr
&&+ \, \phi \, \gamma^{45}(N)
\sum_{n} \gamma_{45}(n) \cr\cr\cr
&=&
\gamma^{\mu 5} \, {\breve p}_\mu - \frac{\phi}{\chi} \,
\gamma^{\mu 4} \, {\breve x}_{\mu} +
\frac{2\phi}{\hbar}\,
\gamma^{\mu\nu} \, {\breve L}_{\mu\nu} + (N-1) \phi \,
\gamma^{45} \, {\breve i}.
\EEqA

In the condensate  all the operators $\gamma^{45}(n)
\gamma_{45}(n')$ attain their minimum eigenvalue $-1$.
Then
\BEq
   (N-1) \, \phi \,
\gamma^{45}\, {\breve i} \; \longrightarrow
\;- \frac{\hbar}{2 \chi}.
\EEq
and the
dynamics becomes
\BEq
\label{eq:DIRACDYNAMICSCONTRACTED}
\tilde{D} \; = \;
\gamma^{\mu5} \, {\breve p}_\mu -
\frac{\phi}{\chi}\,
\gamma^{\mu 4} \, {\breve x}_{\mu} +
\frac{2\phi}{\hbar}\,
\gamma^{\mu\nu} \, {\breve L}_{\mu\nu} - m_\chi c,
\EEq
with rest mass
\BEq
\label{eq:ADJUSTEDMASS}
m_\chi = \frac{\hbar}{2 \chi c}.
\EEq
For sufficently large $N$ this reduces to the usual Dirac
dynamics.

We identify the  mass $m_{\chi}$
with the $N$-independent mass
$m$ of the Dirac equation
for the most massive individual quanta
that the condensate can propagate without melt-down,
on the order of the top quark or Higgs masses:
\BEq
m_{\chi}\sim 10^2 \GeV,\quad \chi\sim 10^{-25} \sec .
\EEq
The universe is $\sim 10^{10}$ years old.
This leads to an upper bound
\BEq
N_{\6M\6a\6x} \sim 10^{41}.
\EEq

This implies that $\chi$ is independent of $N$
as $N\to \infty$ and that $\phi \sim 1/N \to 0$
as $N\to \infty$ even for finite $\chi$. In experiments near
the Higgs energy, $p\sim \hbar /\chi$.  If we also determine
$N$ by setting $x\sim N\chi$ then all  four terms in
(\ref{eq:DIRACDYNAMICSCONTRACTED})  are of the same order of
magnitude.

To estimate experimental effects, however,
we must take gauge transformations into account.
These transform the second term away in the continuum limit.
This refinement of the theory is still in progress.

\section{Conclusions}

   Like classical Newtonian mechanics, the Dirac equation has
a compound (non-semisimple) invariance group. Its variables
break up into three mutually commuting sets: the
space-time-energy-momentum variables ($x^{\mu}$,
$p_{\mu}$), the spin variables $\ga^{\mu}$,
and the imaginary unit $i$.

To unify them we replace the space-time continuum
by an aggregate of $M < \infty$ finite elements,
chronons,
described by spinors with $\sim 2^{M/2}$ components.
Chronons have Clifford-Wilczek statistics,
whose simple operator algebra
is generated by  units $\ga^m$, $m = 1, \dots, M$.
We express
all the variables $x^{\mu}, p_{\mu}, \gamma^{\mu}$ and $i$
as polynomials in the $\ga^m$.
We group the $M=6N$ chronons into $N$ hexads for this
purpose, corresponding to tangent spaces;
the hexad is
the least cell that
suffices for this simplification.
There are three simplifiers
$\chi,
\phi, 1/N$, all approaching 0 in the continuum limit,
   subject to the constraint $\chi \phi \, (N-1) = \hbar/2$ for
all
$N$.

In the continuum limit the Dirac mass becomes infinite.
In our theory, the finite Dirac
masses in nature are consequences of
a finite atomistic quantum space-time structure with
$\chi>0$.

The theory predicts a certain spin-orbit coupling
$\gamma^{\mu\nu} L_{\mu\nu}$ not found in the Standard Model,
and vanishing only in the continuum limit. The experimental
observation of this spin-orbit coupling would further indicate the
existence of a chronon.

In this theory,
the spin we see in nature
is a manifestation of
the (Clifford) statistics of
atomic elements of space-time,
as Brownian motion is of the atomic elements
of matter.
As we improve our theory
we will interpret better other indications of
chronon structure that we already have,
and
as we improve our measuring techniques
we shall meet more such signs.

\section*{ACKNOWLEDGMENTS}
This work was aided by discussions with James Baugh,
Heinrich Saller and Frank Wilczek. It was partially supported
by the M. and H. Ferst Foundation.


\begin{thebibliography}{}

\bibitem{Segal51}
I.E. Segal,
{\it Duke Math. J.} {\bf 18}, 221 (1951)

\bibitem{Inonu52}
E. In\"on\"u and E. P. Wigner, {\it Proc.
Nat. Acad. Sci.} {\bf 39}, 510 (1952)

\bibitem{INONU}
E. In\"on\"u, Contraction of Lie Groups and their
Representations, in F. G\"ursey, ed., {\it Group Theoretical
Consepts and Methods in Elementary Particle Physics}, pp.
391--402 (Gordon and Breach, Science publishers, New York,
1964)

\bibitem{BORN}
M. Born, {\it Rev. Mod. Phys.} {\bf 21}, 463 (1949)

\bibitem{fs}
J. Baugh, D. Finkelstein, A. Galiautdinov, and H.
Saller, {\it J. Math. Phys.} {\bf 42}, 1489 (2001)

\bibitem{FEYNMAN}
R.P. Feynman, private communication.

\bibitem{PENROSE}
R. Penrose,
Angular momentum: an approach
to  combinatorial space-time, in T. Bastin, ed., {\it Quantum
Theory and Beyond}, pp. 151--180 (Cambridge University
Press, Cambridge, 1971)

\bibitem{WEIZSAEKER}
C.F. von Weizs\"acker, {\it Aufbau der Physik} (Hauser,
Munich, 1986)

\bibitem{GALILEO}
G. Galilei, {\it Dialogue concerning the two chief world
systems, ptolemaic and copernican.} Translated by S.
Drake, foreword by A. Einstein. 2d edition  (University
of California Press, Berkeley, 1967)

\bibitem{KEPLERSOMNIUM}
J. Kepler, {\it  Somnium; the dream, or posthumous work on
lunar astronomy.} Translated with a commentary by E.
Rosen (University of Wisconsin Press, Madison, 1967)

\bibitem{FDR69}
D. Finkelstein,
{\em Phys. Rev.} {\bf 184}, 1261 (1969);
{\it ibid.} {\bf D5},  2922 (1972)

\bibitem{DRF96}
D. Finkelstein, {\it Quantum Relativity}
(Springer-Verlag, New York, 1996)

\bibitem{Wilczek82} F. Wilczek and A. Zee,
{\em Phys. Rev.} {\bf D25}, 553 (1982)

\bibitem{NW}
C. Nayak and F. Wilczek, {\it Nucl. Phys.}
{\bf B479}, 529 (1996)

\bibitem{wilczek}
F. Wilczek, hep-th/9806228 [LANL]

\bibitem{FINKELSTEINRODRIGUEZ84}
D.R. Finkelstein and  E. Rodriguez, {\it Int.
J. Theor. Phys.} {\bf 23}, 887 (1984)


\bibitem{FG}
D. Finkelstein and A. Galiautdinov, {\it J.
Math. Phys.} {\bf 42}, 3299 (2001)

\bibitem{DRF99}
D. Finkelstein,
{\em Int. J. Theor. Phys.} {\bf 38}, 447 (1999)

\bibitem{adler}
S. L. Adler, {\it Quaternionic quantum mechanics and quantum
fields} (Oxford University Press, New York, 1995)

\bibitem{Schur1911}
I. Schur, {\it Journal f\"{u}r die reine und
angewandte Mathematik} {\bf 139}, 155 (1911)

\bibitem{ROBSON}
B.A. Robson and D.S. Staudte, {\it J. Phys.} {\bf A29}, 157
(1996); D.S. Staudte, {\it J. Phys.} {\bf A29}, 169 (1996)

\bibitem{LUNSFORD}
D.R. Lunsford, private communication.

\bibitem{SNYDER}
H.S. Snyder,
{\em Phys. Rev.} {\bf 71}, 38 (1947);
{\it ibid.} {\bf 72},  68 (1947)


\bibitem{DOPLICHER}
S. Doplicher, hep-th/0105251 [LANL], and references cited
therein.


\bibitem{STUECKELBERG1938}
E.C.G. St\"uckelberg, {\it Helv. Phys. Acta} {\bf 11}, 299
(1938)

\bibitem{FINKELSTEINJAUCH1959}
D. Finkelstein, J.M. Jauch, and D. Speiser,
Quaternion quantum
mechanics I, II, III.
European Center for Nuclear Research, Geneva,
CERN Reports
59-7, 59-11, 59-17  (1959).
Reprinted in C.A. Hooker, ed., {\em The Logico-Algebraic
Approach to Quantum Mechanics}\/, volume 2, Reidel (1979).

\bibitem{FINKELSTEINJAUCH1962}
D. Finkelstein, J.M. Jauch, S. Schiminovich, and D. Speiser,
{\it J. Math. Phys.} {\bf 4}, 788 (1962)

\bibitem{HIGGS}
P. Higgs, {\it Phys. Lett.} {\bf 12}, 132 (1964); {\it
Phys. Rev. Lett.} {\bf 13}, 508 (1964); {\it Phys. Rev.}
{\bf 145}, 1156 (1966)


\bibitem{DIRAC35}
   P.A.M. Dirac, {\it Ann. Math.} {\bf 36}, 657 (1935)

\end{thebibliography}
\end{document}